\documentclass[prd,showpacs,preprint,nofootinbib]{revtex4}
\usepackage{graphicx,color,amsmath}
\usepackage{bm}
\usepackage{slashed}
\usepackage{extarrows}
\allowdisplaybreaks
\newcommand{\D}{{\rm{d}}}

\begin{document}
\title{Note on the evaluation of one type scalar three-point integral
extracted from the Higgs boson decay}
\author{Jin Zhang\footnote{jinzhang@yxnu.edu.cn}}
\affiliation{School of Physics and Mathematics, Yuxi Normal University,
Yuxi, Yunnan, 653100, China}


\begin{abstract}
Motivated by the Higgs boson decaying to $gg$ at leading order
approximation, the amplitude of scalar one loop three-point
diagram with two different internal masses
are evaluated and fully analytic results are obtained.
The main ingredient of the evaluation
is a integral in which the integrand is product of the reciprocal of the
integral variable and a logarithm whose argument
is a quadratic function of the general form.
Complete analytic results for the two cases that
there is no real root and there are
two different real roots for the logarithm are presented.
Applications of the results in this paper to
radiative decays of the heavy quarks in the Standard Model and
in the singlet vector-like model are discussed.
\end{abstract}


\maketitle
\newpage

\section{introduction}\label{theintroduction}
In the Standard Model (SM) of the particle physics,
the analytic expressions of the scalar one loop three-point
integrals~\cite{npb153.365,prd40.54} play a significant role in
evaluating the amplitudes of triangle diagram mediated processes.
A typical example is deciphering the property of
the Higgs boson~\cite{plb716:1, plb716:30} though
its decaying to $gg$ (or $\gamma\gamma$) and the inverse process, i.e.,
the production of single Higgs boson by gluon fusion~\cite{higgshunter,
kniehl1994, Djouadi:2005gi, spira2017, PDG2024}.
Owing to the couplings of the Higgs boson
to the fermions $g_{Hf\bar{f}}$ and the coupling
of fermions to gluons, the three propagators forming
the triangle take the same mass. In addition to the top quark,
there is considerable mass gap between the Higgs boson and other quarks,
an economic way to compute the amplitude
is taking the limit that $m_{H}\rightarrow \infty$ so that
the evaluation will be greatly simplified and the resulted amplitude
tends to a constant~\cite{Ellis:1975ap, Rizzo:1979mf}.
The advantage of this manner is convenient for evaluation,
but we lose the chance to compare the different contributions from various competing
processes to the amplitude. Thus complete analytic results
are indispensable in order to give a thorough analysis on $H\rightarrow gg$
and its inverse process.
If the high energy approximation is not exploited,
the ultimate amplitude will be expressed as function of
$m_{f}/m_{H}$, where $m_{f}$ and $m_{H}$ are the masses
of the internal fermion and the Higgs boson, respectively.
At the final stage of the evaluation, a integral of the following type
necessarily arises and must be handled carefully
\begin{equation}
I_{1}=\int_{0}^{1}\D x\,
\frac{\ln(ax^{2}-ax+1-i\varepsilon)}{x},  \quad a>0,
\quad \varepsilon\rightarrow 0^{+}  \label{onelooptrianglegeneral}
\end{equation}
Since the argument of the logarithm is quadratic function,
we must distinguish the cases $0<a<4$ from the case
$a>4$~\cite{Djouadi:2005gi, vainshtein1979,
Gunion:1986nh, npb1988279.221, Huang:2011yf, Shifman:2011ri, Marciano:2011gm}.

As a generalization of Eq.(\ref{onelooptrianglegeneral}),
let us consider the following integral
\begin{equation}
I_{2}=\int_{0}^{1}\D x\,
\frac{\ln(ax^{2}-bx+1-i\varepsilon)}{x}, \quad a>0, \quad  a\neq b,
\quad \varepsilon\rightarrow 0^{+}  \label{onelooptypetwo}
\end{equation}
if $a=b$, Eq.(\ref{onelooptypetwo}) reduces to Eq.(\ref{onelooptrianglegeneral}).
After a close inspection on the scalar one loop three point
amplitude we find that the integral given in Eq.(\ref{onelooptypetwo}) will
occur in the evaluation of one type triangle diagram.
Two lines of the triangle have same mass but different from the remaining one,
while two particles in the final state are massless.
Diagrams of the flavor changing neutral currents transition are
of this type. The techniques for analytic evaluating scalar three point
integral has been presented in Ref.~\cite{npb153.365} for decades, since
then there are many works devoting to the evaluation of
scalar one loop three point integral in practical
applications~\cite{kniehl1994, vanOldenborgh:1989wn, Denner:1991kt, Ellis:2007qk, Kniehl:1990mq,
Beenakker:1988bq, Huang:2011yf, npb323.267} and powerful calculating
tools also developed~\cite{Mertig:1990an,
Shtabovenko:2016sxi, Shtabovenko:2020gxv, Kublbeck:1990xc, Hahn:2000kx, Hahn:2000jm}.
However, in these literatures only the final
results are listed while the details of the evaluation are always
omitted. This leads to the situation that it is difficult
to follow the specific evaluating steps, especially for the beginners.
Therefore, in this short paper we will present the thorough steps
for evaluating Eq.(\ref{onelooptypetwo}). Then as a test
of the results, we apply the approach developed in this paper
to evaluate the triangle diagram with two massless external lines.
The special case $a=b$, which is significant for analyzing
$H\rightarrow \gamma\gamma (gg)$, is discussed.

The paper is organized as follows. In section~\ref{evaluationoflogintegral}
we present the analytic result for the integral in Eq.(\ref{onelooptypetwo}).
In section~\ref{evaluationoftriangle} the analytic results of
fig.\ref{twomassive} are obtained. In section~\ref{resultsanddiscus}
the special case $a=b$ are discussed in details and summary of the paper
is presented. Some useful formulas are listed in the appendix.

\section{evaluation of integral with logarithms} \label{evaluationoflogintegral}

We consider the integral of the following type
\begin{equation}
F=\int_{0}^{1}\D x\,\frac{\ln(ax^{2}-bx+1-i\varepsilon)}{x},
\quad a>0        \label{theintegralF}
\end{equation}
Since the argument of the logarithm is quadratic in $x$,
we must distinguish the case $b^{2}-4a<0$ from the case $b^{2}-4a>0$.
Firstly, we consider the case $b^{2}-4a<0$,
in this case the argument of the logarithm
is positive definite thus the $i\varepsilon$ can be safely dropped.
A feasible way to evaluate the integral of Eq.(\ref{theintegralF}) turns out to be
expressing it as
\begin{equation}
F(\alpha)=\int_{0}^{1}\D x\,\frac{1}{x+\alpha}
\ln\frac{ax^2-bx+1}{a\alpha^{2}+b\alpha+1},\quad \alpha>0  \label{Fintegralparameterization}
\end{equation}
It is obvious that
\begin{equation}
F=F(0),  \label{limitingprocess}
\end{equation}
To evaluate $F(\alpha)$, we make the linear transformation
\begin{equation}
mu=x+\alpha,  \label{substitution}
\end{equation}
where the parameter $m$ is to be determined.
Using the method in Ref.~\cite{lewindilog},
Eq.(\ref{Fintegralparameterization}) casts into
\begin{equation}
F(\alpha)=\int_{0}^{(1+\alpha)/m}
\frac{\ln(u^{2}
-2u\cos\theta+1)}{u}\D u
-\int_{0}^{\alpha/m}
\frac{\ln(u^{2}
-2u\cos\theta+1)}{u}\D u,   \label{paraintegralfalpha}
\end{equation}
where
\begin{equation}
m=\sqrt{\frac{a\alpha^{2}+b\alpha+1}{a}},\quad
\cos\theta=\frac{2a\alpha+b}{2\sqrt{a(a\alpha^{2}+b\alpha+1)}}, \label{defofmandtheta}
\end{equation}
Using Eq.(\ref{dilogtwo}) we obtain
\begin{equation}
F(\alpha)=-2{\rm{Li}}_{2}\big(\frac{1+\alpha}{m},\theta\big)
+2{\rm{Li}}_{2}\big(\frac{\alpha}{m},\theta\big),  \label{anlyticresult}
\end{equation}
taking the limit $\alpha\rightarrow0$ we get the desired result
\begin{equation}
F=-2{\rm{Li}}_{2}(\sqrt{a},\theta)={\rm{Re}}\,{\rm{Li}}_{2}(\sqrt{a}e^{i\theta}), \label{finalreultsoffirstcase}
\end{equation}
with the parameter $\theta$ given by
\begin{equation}
\theta=\arccos\frac{b}{2\sqrt{a}}.  \label{defoftheta}
\end{equation}

Then we consider the case $b^{2}-4a>0$. Now there are
two zeros of the logarithm in the range $[0,\,1]$,
the $i\varepsilon$ prescription must be retained appropriately.
By employing integration by parts, we find
\begin{equation}
F=-\int_{0}^{1}\frac{(2ax-b)\ln x}{a(x-x_{1})(x-x_{2})}\,\D x, \label{firstkindcasetwo}
\end{equation}
where $x_{1}$ and $x_{2}$ are the two roots of the argument of the logarithm
\begin{eqnarray}
x_{1}&=&x_{+}+i\varepsilon,\quad \quad
x_{+}=\frac{b+\sqrt{b^{2}-4a}}{2a},\nonumber\\
x_{2}&=&x_{-}-i\varepsilon,\quad\quad
x_{-}=\frac{b-\sqrt{b^{2}-4a}}{2a}, \label{tworootsoffirskindcasetwo}
\end{eqnarray}
By making use the partial fraction expansion, we arrive the following result
\begin{eqnarray}
F&=&\frac{1}{x_{1}-x_{2}}(\frac{b}{a}-2x_{1})\int_{0}^{1}\frac{\ln x}{x-x_{1}}\,\D x
+\frac{1}{x_{1}-x_{2}}(2x_{2}-\frac{b}{a})
\int_{0}^{1}\frac{\ln x}{x-x_{2}}\,\D x\nonumber\\
&=&\frac{1}{x_{1}-x_{2}}\Big\{(\frac{b}{a}-2x_{1})
{\rm{Li}}_{2}\Big[\frac{1}{x_{+}}-i\varepsilon\, {\rm{sgn}}(x_{+})\Big]
-(\frac{b}{a}-2x_{2}){\rm{Li}}_{2}
\Big[\frac{1}{x_{-}}+i\varepsilon\, {\rm{sgn}}(x_{-})\Big]\Big\}.   \label{resultoffirstkindcasetwo}
\end{eqnarray}
with the function ${\rm{sgn}}(x)$ defined in Eq.(\ref{signfunction}).

\section{The massive triangle with two massless external lines}\label{evaluationoftriangle}

\subsection{case one: $a\neq b$} \label{caseofunequalmass}

\begin{figure}
\begin{center}
\includegraphics[scale=0.60]{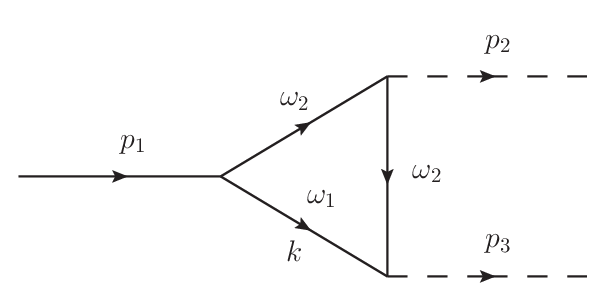}
\caption{Massive triangle with two massless external lines. The solid lines and dashed
lines denote massive and massless particles, respectively.}\label{twomassive}
\end{center}
\end{figure}

To start with, we write down the amplitude corresponding to fig.\ref{twomassive}
\begin{equation}
I=\int\frac{\D^{4}k}{(2\pi)^{4}}
\frac{1}{A_{1}A_{2}A_{3}}, \label{threemassivetriangle}
\end{equation}
where for brevity we set the couplings of the three vertexes
to be unit one, the three denominators are defined by
\begin{eqnarray}
A_{1}&=&k^{2}-\omega_{1}^{2}+i\varepsilon\nonumber\\
A_{2}&=&(p_{1}+k)^{2}-\omega_{2}^{2}+i\varepsilon\nonumber\\
A_{3}&=&(p_{1}-p_{2}+k)^{2}-\omega_{2}^{2}+i\varepsilon,    \label{definitionofthreedenoms}
\end{eqnarray}
and $\varepsilon$ is real positive infinitesimal, the three external momentum
satisfy
\begin{equation}
p_{1}^{2}=m_{1}^{2},\quad p_{2}^{2}=p_{3}^{2}=0.  \label{onshellmomentum}
\end{equation}
Using the Feynman's trick, Eq.(\ref{threemassivetriangle}) can be written as
\begin{equation}
I=\int\frac{\D^{4}k}{(2\pi)^{4}}\int\D x\D y\D z
\frac{2!\,\delta(1-x-y-z)}{[D(x,y,z)]^{3}}, \label{feynmantrick}
\end{equation}
where
\begin{equation}
D(x,y,z)=x(k^{2}-\omega_{1}^{2}+i\varepsilon)+y[(p_{1}+k)^{2}-\omega_{2}^{2}+i\varepsilon]
+z[(p_{1}-p_{2}+k)^{2}-\omega_{2}^{2}+i\varepsilon].  \label{feynmantrickdenom}
\end{equation}
Since the amplitude given by Eq.(\ref{threemassivetriangle})
is both ultraviolet and infrared finite, thus regularization is unnecessary,
the evaluation can be carried out in the four-dimensional space-time.
We first perform the integral over $k$ and $z$, obtaining
\begin{equation}
I=-\frac{i}{16\pi^{2}}\int_{0}^{1}\D x\int_{0}^{1-x}\D y\,\frac{1}{-yxm_{1}^{2}
+x(\omega_{1}^{2}-\omega_{2}^{2})+\omega_{2}^{2}-i\varepsilon},   \label{momentumandzinted}
\end{equation}
The integral over $y$ in Eq.(\ref{momentumandzinted}) is trivial,
combining with Eq.(\ref{definitionofdilog}),
we arrive at the following intermediate result
\begin{eqnarray}
I&=&\frac{i}{16\pi^{2}m_{1}^{2}}\int_{0}^{1}\D x\frac{1}{x}
\Big[\ln\Big(\frac{m_{1}^{2}}{\omega_{2}^{2}}x^{2}
-\frac{m_{1}^{2}-\omega_{1}^{2}+\omega_{2}^{2}}{\omega_{2}^{2}}x
+1-i\varepsilon\Big)
-\ln\Big(\frac{\omega_{1}^{2}-\omega_{2}^{2}}{\omega_{2}^{2}}x+1-i\varepsilon\Big)\Big]\nonumber\\
&=&\frac{i}{16\pi^{2}m_{1}^{2}}
\Big[{\rm{Li}}_{2}\Big(1-\frac{\omega_{1}^{2}}{\omega_{2}^{2}}\Big)
+\int_{0}^{1}\D x\frac{1}{x}
\ln\Big(\frac{m_{1}^{2}}{\omega_{2}^{2}}x^{2}
-\frac{m_{1}^{2}-\omega_{1}^{2}+\omega_{2}^{2}}{\omega_{2}^{2}}x
+1-i\varepsilon\Big)\Big].  \label{integraloverxleft}
\end{eqnarray}
In order to apply the results in Section~\ref{evaluationoflogintegral} to
evaluate the remaining integral in Eq.(\ref{integraloverxleft}), we identify
\begin{equation}
a=\frac{m_{1}^{2}}{\omega_{2}^{2}},\quad\quad
b=\frac{m_{1}^{2}-\omega_{1}^{2}+\omega_{2}^{2}}{\omega_{2}^{2}},   \label{coefficientsaandb}
\end{equation}
If $b^{2}-4a<0$, this implies that
$\lambda(m_{1}^{2},\omega_{1}^{2},\omega_{2}^{2})<0$,
where $\lambda(x,y,z)$ is the well-known K$\ddot{a}$llen function
\begin{equation}
\lambda(x,y,z)=x^{2}+y^{2}+z^{2}-2xy-2xz-2yz,   \label{kallenfunction}
\end{equation}
by making use Eq.(\ref{finalreultsoffirstcase}) we immediately arrive the final result
\begin{eqnarray}
I
&=&\frac{i}{16\pi^{2}m_{1}^{2}}
\Big[{\rm{Li}}_{2}\Big(1-\frac{\omega_{1}^{2}}{\omega_{2}^{2}}\Big)
-2{\rm{Re}}\,{\rm{Li}}_{2}(\frac{m_{1}}{\omega_{2}}e^{i\theta})\Big]  \label{finalresultscaseone}
\end{eqnarray}
where
\begin{equation}
\theta=\arccos\frac{m_{1}^{2}-\omega_{1}^{2}+\omega_{2}^{2}}{2m_{1}\omega_{2}}.  \label{definitionoftheta}
\end{equation}

If $\lambda(m_{1}^{2},\omega_{1}^{2},\omega_{2}^{2})>0$,
by employing Eq.(\ref{resultoffirstkindcasetwo}) we get
\begin{equation}
I=\frac{i}{16\pi^{2}m_{1}^{2}}\Big\{\frac{1}{2}\,
{\rm{Li}}_{2}(1-\frac{\omega_{1}^{2}}{\omega_{2}^{2}})
-{\rm{Li}}_{2}[\frac{1}{x_{+}}-i\varepsilon\, {\rm{sgn}}(x_{+})]
-{\rm{Li}}_{2}
[\frac{1}{x_{-}}+i\varepsilon\, {\rm{sgn}}(x_{-})]\Big\},  \label{finalresultofIcasetwo}
\end{equation}
where
\begin{eqnarray}
x_{+}&=&\frac{(m_{1}^{2}-\omega_{1}^{2}+\omega_{2}^{2})
+\lambda^{1/2}(m_{1}^{2},\omega_{1}^{2},\omega_{2}^{2})}{2m_{1}^{2}},   \nonumber\\
x_{-}&=&\frac{(m_{1}^{2}-\omega_{1}^{2}+\omega_{2}^{2})
-\lambda^{1/2}(m_{1}^{2},\omega_{1}^{2},\omega_{2}^{2})}{2m_{1}^{2}},  \label{specificxplusandminus}
\end{eqnarray}
are the two roots of the argument of the logarithm in the integrand.

\subsection{case two: $a=b$} \label{caseofequalmass}
It is instructive to consider the special case that the
coefficients in Eq.(\ref{theintegralF}) satisfy the condition $a=b$.
This is the integral indispensable in evaluation of $H\rightarrow gg$ decay at
one loop approximation. Supposing the mass of each propagator of the triangle is $\omega_{1}$,
setting
\begin{equation}
a=b=\frac{m_{1}^{2}}{\omega_{1}^{2}},    \label{ratioofmasssquared}
\end{equation}
hence Eq.(\ref{integraloverxleft}) reduces to
\begin{equation}
I=\frac{i}{16\pi^{2}m_{1}^{2}}\int_{0}^{1}
\D x \frac{\ln(ax^{2}-ax+1-i\varepsilon)}{x},  \label{simplifiedcasetwo}
\end{equation}
Firstly consider the case $0<a<4$, i.e., $0<m_{1}<2\omega_{1}$.
To be conveniently get the correct results we must trace back to
Eq.(\ref{Fintegralparameterization}) other than
simply setting $a=b$ in Eq.(\ref{finalresultscaseone}),
otherwise it will be in trouble to obtain the
final result since the analytic properties
of the dilogarithm function of complex argument are nontrivial.
An advisable way is that we may write it as parameter integral
\begin{eqnarray}
I&=&\frac{ia}{16\pi^{2}m_{1}^{2}}\int_{0}^{1}\D x\int_{0}^{1}
\D z\frac{x-1}{1+zax(x-1)},  \label{equalcoefficientsone}
\end{eqnarray}
Changing the integration order and letting
\begin{equation}
x=u+\frac{1}{2},   \label{subsitutiontou}
\end{equation}
Eq.(\ref{equalcoefficientsone}) cast into
\begin{eqnarray}
I&=&\frac{-ia}{16\pi^{2}m_{1}^{2}}\int_{0}^{1}\D z\int_{0}^{1/2}
\D u\frac{1}{1-\frac{1}{4}za+zau^{2}}\nonumber\\
&=&-\frac{i}{16\pi^{2}m_{1}^{2}}\int_{0}^{1}\D z
\frac{\sqrt{a}}{\sqrt{z}\sqrt{1-\frac{za}{4}}}\arcsin\Big(\frac{\sqrt{za}}{2}\Big)\nonumber\\
&=&-\frac{i}{8\pi^{2}m_{1}^{2}}\Big(\arcsin\frac{\sqrt{a}}{2}\Big)^{2},   \label{uintegrationcompleted}
\end{eqnarray}
where in deriving the above equation the identity~\cite{integraltable2014}
\begin{equation}
\arctan x=\arcsin\frac{x}{\sqrt{1+x^{2}}},   \label{arctantoarcsin}
\end{equation}
has been used. Using the fact that $a=m_{1}^{2}/\omega_{1}^{2}$, we get the well-known function
appeared in one loop evaluation of $H\rightarrow gg$
\begin{equation}
I=-\frac{i}{8\pi^{2}m_{1}^{2}}
\Big(\arcsin\frac{m_{1}}{2\omega_{1}}\Big)^{2},
\quad\quad    0<m_{1}<2\omega_{1}. \label{finalresultofspecialcase}
\end{equation}

Then consider the case of $a>4$, i.e., $m_{1}>2\omega_{1}$.
In this case there are two zeros for the argument of the
logarithm. From Eq.(\ref{resultoffirstkindcasetwo}) we get
\begin{equation}
I=\frac{i}{16\pi^{2}m_{1}^{2}}\Big[-{\rm{Li}}_{2}(\frac{1}{x_{-}}+i\varepsilon)
-{\rm{Li}}_{2}(\frac{1}{x_{+}}-i\varepsilon)\Big], \label{specialcaseofaequaltob}
\end{equation}
$x_{+}$ and $x_{-}$ are given by
\begin{equation}
x_{+}=\frac{1}{2}\Big(1+\sqrt{1-\frac{4}{a}}\,\Big),\quad \quad
x_{-}=\frac{1}{2}\Big(1-\sqrt{1-\frac{4}{a}}\,\Big), \label{simplifiedxplusandminus}
\end{equation}
Since both $x_{+}$ and $x_{-}$ are less than $1$,
in order to simplify the Eq.(\ref{specialcaseofaequaltob}),
defining
\begin{equation}
\alpha=\frac{1}{x_{-}}>1, \quad
\frac{1}{x_{+}}=\frac{1}{1-x_{-}}=\frac{\alpha}{\alpha-1}>1,   \label{alphaandxplusminus}
\end{equation}
Then Eq.(\ref{specialcaseofaequaltob}) can be written as
\begin{equation}
I=\frac{i}{16\pi^{2}m_{1}^{2}}\Big[-{\rm{Li}}_{2}(\alpha+i\varepsilon)
-{\rm{Li}}_{2}(\frac{\alpha}{\alpha-1}-i\varepsilon)\Big],   \label{alphasunbstituted}
\end{equation}
Combining Eq.(\ref{analyticacontofdilog}) we obtain
\begin{eqnarray}
I&=&\frac{i}{16\pi^{2}m_{1}^{2}}\Big\{-\Big[{\rm{Re}}\,{\rm{Li}}_{2}(\alpha)+i\pi\ln\alpha\Big]
-\Big[{\rm{Re}}\,{\rm{Li}}_{2}(\frac{\alpha}{\alpha-1})
-i\pi\ln\frac{\alpha}{\alpha-1}\Big]\nonumber\\
&=&\frac{1}{m_{1}^{2}}\Big\{-{\rm{Re}}
\Big[{\rm{Li}}_{2}(\alpha)+{\rm{Li}}_{2}(\frac{\alpha}{\alpha-1})\Big]
-i\pi\ln(\alpha-1)\Big\},  \label{intermmidateresult}
\end{eqnarray}
In considering the property of the dilogarithm
\begin{equation}
{\rm{Re}}\Big[\,{\rm{Li}}_{2}(x)+{\rm{Li}}_{2}(\frac{x}{x-1})\Big]
=\frac{\pi^{2}}{2}-\frac{\ln^{2}(x-1)}{2}, \quad  x>1   \label{dilogproperty}
\end{equation}
which leads to
\begin{equation}
I=\frac{i}{16\pi^{2}m_{1}^{2}}
\Big[-\frac{\pi^{2}}{2}+\frac{\ln^{2}(\alpha-1)}{2}
-i\pi\ln(\alpha-1)\Big],  \label{realandimaginary}
\end{equation}
Substituting Eq.(\ref{alphaandxplusminus})
into Eq.(\ref{realandimaginary}) and noticing that
\begin{equation}
\alpha-1=\frac{x_{+}}{x_{-}}, \label{alphafromxplusandminus}
\end{equation}
we get
\begin{equation}
I=\frac{i}{16\pi^{2}m_{1}^{2}}\Big(-\frac{\pi^{2}}{2}+\frac{1}{2}
\ln^{2}\frac{x_{+}}{x_{-}}-i\pi\ln\frac{x_{+}}{x_{-}}\Big),  \label{xplusandminusinto}
\end{equation}
Plugging Eq.(\ref{simplifiedxplusandminus}) into Eq.(\ref{xplusandminusinto})
we arrive at the well-known function in the one loop evaluation of $H\rightarrow gg$ decay
\begin{equation}
I=-\frac{i}{32\pi^{2}m_{1}^{2}}\Big(\pi+i\ln\frac{1+\sqrt{1-\frac{4\omega_{1}^{2}}{m_{1}^{2}}}}
{1-\sqrt{1-\frac{4\omega_{1}^{2}}{m_{1}^{2}}}}\Big)^{2},\quad
m_{1}>2\omega_{1}.       \label{finalresultaequaltob}
\end{equation}

\section{discussions and summary}\label{resultsanddiscus}

\begin{figure}
\begin{center}
\includegraphics[scale=0.60]{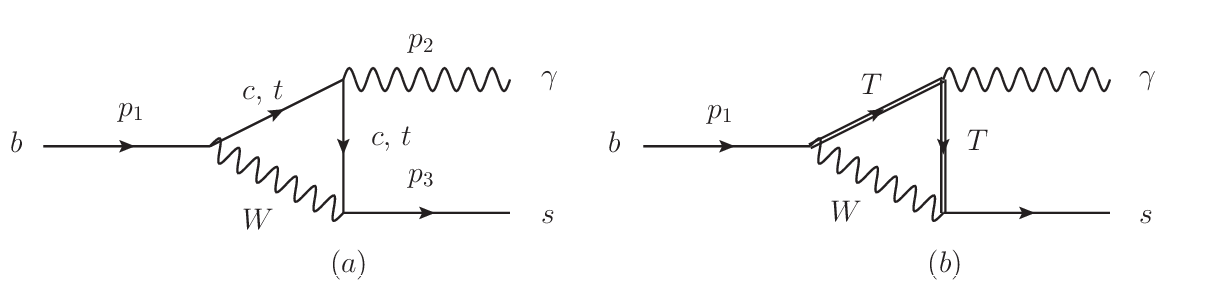}
\caption{$b\rightarrow s\gamma$ decay in the SM and
singlet vector-like quark model. The double lines
denote the singlet vector-like top quark. }\label{heavyquarkdecay}
\end{center}
\end{figure}

\begin{figure}
\begin{center}
\includegraphics[scale=0.60]{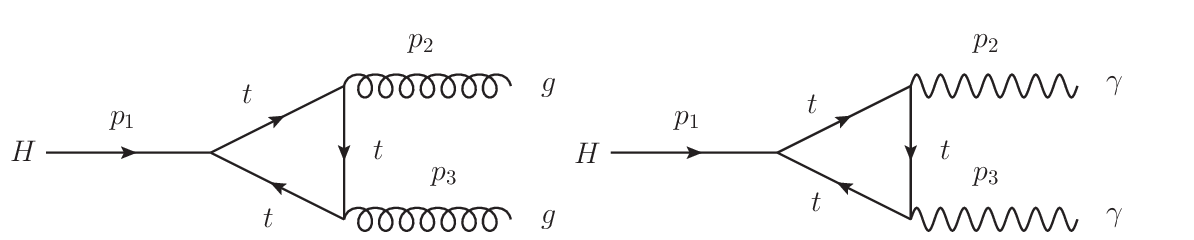}
\caption{The decay of the Higgs boson to $gg$ and to $\gamma\gamma$ in the SM.}\label{decayofthehiggs}
\end{center}
\end{figure}

The analytic expressions derived in section \ref{evaluationoftriangle}
are based on the pure scalar triangle diagram, but it does not
mean that the results only can apply to processes
involving scalar particles. In fact, the results
can be used to evaluate the amplitudes
of any triangle diagram mediated
processes with two massless final states, and the
three propagators may be fermions or/and bosons.
Since after the appropriate tensor structure extracted,
it always needs to deal with scalar three point integrals.
To illustrate this we list some applying circumstances.

The first one is evaluating the amplitude of the
radiative decays of the heavy quarks in the SM.
To be specific we take $b\rightarrow s\,\gamma$ as an example,
which is shown in diagram a of fig.\ref{heavyquarkdecay}. Since the masses
of the up and the strange quark can be neglected compared with
the bottom quark, the mass parameters of the amplitude for this decay are
\begin{equation}
m_{1}=m_{b},\quad \omega_{1}=m_{W},\quad
\omega_{2}=m_{q}(q=c,\,t),      \label{massparamertersone}
\end{equation}
If we extend the SM by including a singlet
vector-like top quark~\cite{delAguila:1985mk, Branco:1986my}
(which is also called as top partner),
then there is new contribution produced by the top partner
to the amplitude of $b\rightarrow s\gamma$ decay
as shown in the diagram b of fig.\ref{heavyquarkdecay}.
In this case, to obtain the results for the scalar integral,
we just need do the following substitution
\begin{equation}
m_{1}=m_{b},\quad \omega_{1}=m_{W},\quad
\omega_{2}=m_{T}, \label{massparamerterstwo}
\end{equation}
where $m_{T}$ denotes the mass of the singlet top partner.

The above two examples match to the scenario that
the masses of the triangle are different, if the masses
of the three propagators are taking the same value, this is the
case analyzed in the second part of section \ref{evaluationoftriangle}.
The decays $H\rightarrow gg, \gamma\gamma$ (and its inverse processes)
are fallen into this category which is presented in fig.\ref{decayofthehiggs}.
In this case the three internal
lines are fermions and the masses of them are identical,
the mass parameters are
\begin{equation}
m_{1}=m_{H},\quad \omega_{1}=\omega_{2}=m_{t}.   \label{massparamertersthree}
\end{equation}
where only the contribution of top quark is taken into accounted.
Then with results presented in the second part of
section \ref{evaluationoftriangle}, it is easily to obtain the
desired analytic results.

In summary, the amplitude of the scalar one loop three-point
diagram in which the three propagators are assigned two different masses is evaluated.
A general type integral is extracted and analytic results are obtained.
The possible applications of the results are discussed.
If each propagator of the triangle taking the same mass,
we find that the general results will simplify to the functions obtained in the
lowest order evaluation of $H\rightarrow gg$ decay.
The results and techniques presented in this paper can be applied also
to decaying processes which are mediated by one loop triangle diagram.

\section*{Acknowledgements}
The author thanks H. E. Haber for reminding the work in
Ref.~\cite{npb323.267} and the updated version in his
webpage.

\begin{appendix}

\section{The dilogarithm function}\label{propertiesofthedilog}

We list some necessary formulas in our evaluation.
The dilogarithm is defined as~\cite{lewindilog}
\begin{equation}
{\rm{Li}}_{2}(x)=\sum_{k=1}^{+\infty}\frac{x^{k}}{k^{2}}
=-\int_{0}^{1}\frac{\ln(1-xt)}{t}\,\D t, \quad  |x|<1\label{definitionofdilog}
\end{equation}
There is a branch cut from $1$ to $\infty$, for $\varepsilon \rightarrow 0$
\begin{equation}
{\rm{Li}}_{2}(x+i\varepsilon)
={\rm{Re}}\,{\rm{Li}}_{2}(x)
+i\pi\,{\rm{sgn}}(\varepsilon)\Theta(x-1)\ln x,   \label{analyticacontofdilog}
\end{equation}
where $\Theta(x)$ and ${\rm{sgn}}(x)$ are the step and sign function, respectively.
They are defined as
\begin{equation}
\Theta(x)=
\begin{cases}
1   \quad\quad  x>0,\nonumber\\
0  \quad\quad x<0,
\end{cases} \quad
{\rm{sgn}}(x)=
\begin{cases}
1   \quad\quad\quad  x>0,\nonumber\\
-1  \quad\quad x<0,
\end{cases}   \label{signfunction}
\end{equation}
Another useful formula is
\begin{equation}
{\rm{Li}}_{2}(x,\theta)={\rm{Re}}\,{\rm{Li}}_{2}(xe^{i\theta})
=-\frac{1}{2}\int_{0}^{x}\frac{\ln(t^{2}-2t\cos\theta+1)}{t}\,\D t.  \label{dilogtwo}
\end{equation}

\end{appendix}

\end{document}